# Commutation Relations for Double Tensors of Two Equivalent D Electrons

Chin-Sheng Wu
Center for General Education
Yuan Ze University, Ne-Li, Taiwan

**Abstract.** We apply the Clebsch-Gordan and Racah coefficients to calculate the double tensors for two equivalent d electrons. We also obtain the commutation relations for these double tensors and choose certain quantum numbers, which produce a subgroup. From the root vectors of the commutation relations, we identify them with Lie algebra $B_2$. Once we have the correct Lie algebra, it's feasible to use the Wigner-Eckart theorem to find matrix elements for transition states among atomic spectra or nuclear shell models.

## I. Introduction

Group theory has been applied extensively in atomic spectroscopy, nuclear shell theory and elementary particles. The use of tensor operators[1-4] with the second quantization puts many features of these physical fields in better perspective.
In atomic spectroscopy, the creation and annihilation operators provide a way to form a variety of tensor operators, which connect different configurations in two folds. One is the transitions produced by tensor operators acting on eigenfunctions for each shell. The other is the group structures for certain quantum numbers, which form some subgroups. The spectra are therefore classified. In Section II, we choose the two equivalent d electrons as an example to realize the Racah's double tensors method.

## II. Calculation and Results

The usual rules of angular momentum theory define the coupled double tensor as following formula:

$$(a^+ a)^{\sigma\ k}_{\pi\ q} = \sum_{\xi\ \eta} \langle sm_s sm_s' | ss\sigma\pi \rangle \langle lm_l lm_l' | llkq \rangle a^+_\xi a_\eta$$

$$\xi = (m_s, m_l)$$

$$\eta = (m_s', m_l')$$

The commutation relations of double tensors for two equivalent d electrons can be calculated through the following formula:

$$\left[\left(a^+a\right)^\sigma_\pi{}^k_q, \left(a^+a\right)^{\sigma'}_{\pi'}{}^{k'}_{q'}\right] = \sum_{\sigma'',k'',\pi'',q''}\{(2\sigma+1)(2\sigma'+1)(2k+1)(2k'+1)\}^{\frac{1}{2}}\{(-1)^{\sigma+k+\sigma'+k'} - (-1)^{\sigma''+k''}\}$$

$$\langle\sigma\pi\sigma'\pi'|\sigma\sigma'\sigma''\pi''\rangle\langle kqk'q'|kk'k''q''\rangle W(\sigma\sigma'ss;\sigma''s)W(kk'll;k''l)\left(a^+a\right)^{\sigma''}_{\pi''}{}^{k''}_{q''},$$

.

where the Clebsch-Gordan and Racah coefficients are used.

The term $(-1)^{\sigma+k+\sigma'+k'} - (-1)^{\sigma''+k''}$ in the commutation formula suggests that odd quantum numbers of angular momentum give the odd $k''$ if we choose the addition of spin as zero.

$$\left[\left(a^+a\right)^0_0{}^1_1, \left(a^+a\right)^0_0{}^1_{-1}\right] = 0.2236\left(a^+a\right)^0_0{}^1_0 = \frac{1}{2\sqrt{5}}\left(a^+a\right)^0_0{}^1_0$$

$$\left[\left(a^+a\right)^0_0{}^1_1, \left(a^+a\right)^0_0{}^1_0\right] = 0.2236\left(a^+a\right)^0_0{}^1_1 = \frac{1}{2\sqrt{5}}\left(a^+a\right)^0_0{}^1_1$$

$$\left[\left(a^+a\right)^0_0{}^1_{-1}, \left(a^+a\right)^0_0{}^1_0\right] = -0.2236\left(a^+a\right)^0_0{}^1_{-1} = \frac{1}{2\sqrt{5}}\left(a^+a\right)^0_0{}^1_{-1}$$

$$\left[\left(a^+a\right)^0_0{}^3_{-3}, \left(a^+a\right)^0_0{}^3_3\right] = -\frac{3}{2\sqrt{5}}\left(a^+a\right)^0_0{}^1_0 + \frac{1}{2\sqrt{5}}\left(a^+a\right)^0_0{}^3_0$$

$$\left[\left(a^+a\right)^0_0{}^3_{-2}, \left(a^+a\right)^0_0{}^3_2\right] = \frac{1}{\sqrt{5}}\left(a^+a\right)^0_0{}^1_0 + \frac{1}{2\sqrt{5}}\left(a^+a\right)^0_0{}^3_0$$

$$\left[\left(a^+a\right)^0_0{}^3_{-1}, \left(a^+a\right)^0_0{}^3_1\right] = -\frac{1}{2\sqrt{5}}\left(a^+a\right)^0_0{}^1_0 + \frac{1}{2\sqrt{5}}\left(a^+a\right)^0_0{}^3_0$$

$$\left[\left(a^+a\right)^0_0{}^1_0, \left(a^+a\right)^0_0{}^3_q\right] = -\frac{q}{2\sqrt{5}}\left(a^+a\right)^0_0{}^3_q$$

$$\left[\left(a^+a\right)^0_0{}^1_{\pm 1}, \left(a^+a\right)^0_0{}^3_q\right] = \mp\frac{1}{2\sqrt{5}}\sqrt{(3\mp q)(3\pm q+1)}\left(a^+a\right)^0_0{}^3_{q\pm 1}$$

$$\left[\left(a^+a\right)^0_0{}^1_{-1}, \left(a^+a\right)^0_0{}^3_{-2}\right] = -\frac{\sqrt{3}}{2\sqrt{5}}\left(a^+a\right)^0_0{}^3_{-3}$$

$$\left[(a^+a)^0{}_0{}^1{}_{-1},(a^+a)^0{}_0{}^3{}_{-1}\right]=-\frac{1}{2}(a^+a)^0{}_0{}^3{}_{-2}$$

$$\left[(a^+a)^0{}_0{}^3{}_{-3},(a^+a)^0{}_0{}^3{}_{0}\right]=\frac{1}{2\sqrt{5}}(a^+a)^0{}_0{}^3{}_{-3}$$

$$\left[(a^+a)^0{}_0{}^3{}_{-3},(a^+a)^0{}_0{}^3{}_{1}\right]=\frac{1}{2\sqrt{15}}(a^+a)^0{}_0{}^3{}_{-2}$$

$$\left[(a^+a)^0{}_0{}^3{}_{-3},(a^+a)^0{}_0{}^3{}_{2}\right]=-\frac{3}{2\sqrt{5}}(a^+a)^0{}_0{}^1{}_{-1}+\frac{1}{2\sqrt{15}}(a^+a)^0{}_0{}^3{}_{-1}$$

$$\left[(a^+a)^0{}_0{}^3{}_{-2},(a^+a)^0{}_0{}^3{}_{1}\right]=\frac{1}{2}(a^+a)^0{}_0{}^3{}_{-1}$$

$$\left[(a^+a)^0{}_0{}^3{}_{-2},(a^+a)^0{}_0{}^3{}_{0}\right]=\frac{1}{2\sqrt{5}}(a^+a)^0{}_0{}^3{}_{-2}$$

$$\left[(a^+a)^0{}_0{}^3{}_{-2},(a^+a)^0{}_0{}^3{}_{3}\right]=-\frac{\sqrt{3}}{2\sqrt{5}}(a^+a)^0{}_0{}^1{}_{1}+\frac{1}{2\sqrt{15}}(a^+a)^0{}_0{}^3{}_{1}$$

$$\left[(a^+a)^0{}_0{}^3{}_{-1},(a^+a)^0{}_0{}^3{}_{0}\right]=-\frac{\sqrt{3}}{\sqrt{10}}(a^+a)^0{}_0{}^1{}_{-1}-\frac{1}{2\sqrt{5}}(a^+a)^0{}_0{}^3{}_{-1}$$

$$\left[(a^+a)^0{}_0{}^3{}_{-1},(a^+a)^0{}_0{}^3{}_{2}\right]=\frac{1}{2}(a^+a)^0{}_0{}^1{}_{1}$$

$$\left[(a^+a)^0{}_0{}^3{}_{-1},(a^+a)^0{}_0{}^3{}_{3}\right]=\frac{1}{2\sqrt{15}}(a^+a)^0{}_0{}^1{}_{2}$$

$$\left[(a^+a)^0{}_0{}^3{}_{0},(a^+a)^0{}_0{}^3{}_{1}\right]=-\frac{\sqrt{3}}{\sqrt{10}}(a^+a)^0{}_0{}^1{}_{1}-\frac{1}{2\sqrt{5}}(a^+a)^0{}_0{}^3{}_{-1}$$

$$\left[(a^+a)^0{}_0{}^3{}_{0},(a^+a)^0{}_0{}^3{}_{2}\right]=-\frac{1}{2\sqrt{5}}(a^+a)^0{}_0{}^1{}_{2}$$

$$\left[(a^+a)^0{}_0{}^3{}_{0},(a^+a)^0{}_0{}^3{}_{3}\right]=\frac{1}{2\sqrt{5}}(a^+a)^0{}_0{}^1{}_{3}$$

$$\left[(a^+a)^{0\ 3}_{0\ 1},(a^+a)^{0\ 3}_{0\ 2}\right]=-\frac{1}{2\sqrt{15}}(a^+a)^{0\ 1}_{0\ 3}$$

Those commutations are omitted if they are zero or can be found by switching the tensors in the above equations. In order to match $B_2$ Lie algebra, we define the following related tensor operators

$b=1/0.2236=2\sqrt{5}$,

$$J_0 = b\,(a^+a)^{0\ 1}_{0\ 0},$$

$$J_1 = b\,(a^+a)^{0\ 1}_{0\ 1},$$

$$J_{-1} = b\,(a^+a)^{0\ 1}_{0\ -1}.$$

$$F_0 = b(a^+a)^{0\ 3}_{0\ 0}$$

$$F_1 = b(a^+a)^{0\ 3}_{0\ 1}$$

$$F_{-1} = b(a^+a)^{0\ 3}_{0\ -1}$$

$$F_2 = b(0.6(a^+a)^{0\ 3}_{0\ 2}+0.4(a^+a)^{0\ 3}_{0\ 3})$$
$$F_{-2} = b(0.6(a^+a)^{0\ 3}_{0\ -2}+0.4(a^+a)^{0\ 3}_{0\ -3})$$
$$F_3 = b(0.4(a^+a)^{0\ 3}_{0\ 2}+0.6(a^+a)^{0\ 3}_{0\ 3})$$
$$F_{-3} = b(0.4(a^+a)^{0\ 3}_{0\ -2}+0.6(a^+a)^{0\ 3}_{0\ -3})$$

Now the above new double tensors form the basis vectors of $B_2$ Lie algebra as in Fig. 1. There are many other choices for $B_2$ Lie algebra.

$$[J_0, J_1] = J_1$$
$$[J_0, J_{-1}] = J_{-1}$$
$$[J_1, J_{-1}] = J_0$$

$$[F_1, F_{-1}] = \sqrt{2}J_0 + \sqrt{2}F_0$$

$$[F_2, F_{-2}] = F_0$$

$$[F_3, F_{-3}] = -\sqrt{2}J_0 + \sqrt{2}F_0$$

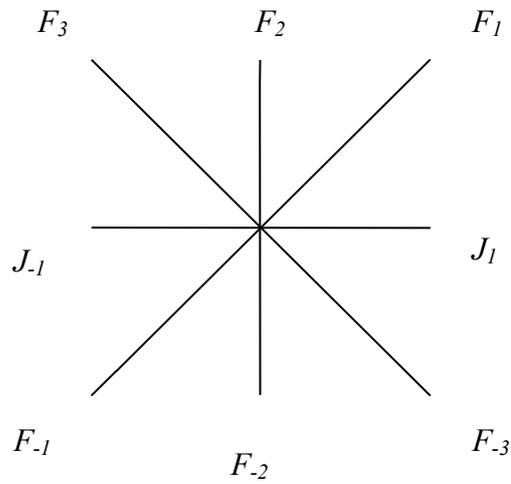

Fig. 1. The root vectors of $B_2$ Lie algebra.